\documentclass[12pt]{article}
\usepackage{latexsym}
\usepackage{pgf}
\newcommand{\beq}{\begin{equation}}
\newcommand{\eeq}{\end{equation}}
\newcommand{\beqs}{\begin{eqnarray}}
\newcommand{\eeqs}{\end{eqnarray}}
\newcommand{\tr}{\mathrm{tr}}

\begin{document}
\begin{titlepage}
~ {\hfill CCNY-HEP-06/1}
\vskip 1.2cm
\begin{center}
{\LARGE BPS Operators in $\cal{N}$=4 SYM:}\\{\Large
Calogero Models and 2D Fermions}\\
\vskip 1.2cm
{\Large Abhishek Agarwal} \, and \, {\Large Alexios P. Polychronakos}\\
\vskip 1.3cm {\Large Physics Department}\\
{\large City College of the CUNY}\\
{\large New York, NY 10031}\\
abhishek,alexios@sci.ccny.cuny.edu
\vskip 0.7cm
February 3, 2006
\end{center}
\vspace{2cm}
\begin{abstract}

A connection between the gauge fixed dynamics of protected operators
in superconformal Yang-Mills theory in four dimensions and Calogero
systems is established. This connection generalizes the free Fermion
description of the chiral primary operators of the gauge theory
formed out of a single complex scalar to more general operators. In
particular, a detailed analysis of protected operators charged under
an $su(1|1)$ $\in $ $psu(2,2|4)$ is carried out and a class of
operators is identified, whose dynamics is described by the rational
super-Calogero model. These results are generalized to arbitrary BPS
operators  charged under  an $su(2|3)$ of the superconformal
algebra. Analysis of the non-local symmetries of the super-Calogero
model is also carried out, and it is shown that symmetry for  a
large class of protected operators is a contraction of the
corresponding Yangian algebra  to a loop algebra.
\end{abstract}
\end{titlepage}

\section{Introduction}A substantial number of problems related to the study of maximally
supersymmetric  Yang-Mill theory on $R\times S^3$ can be translated
into the study of hamiltonian multi-matrix models. Perhaps the most
striking success of this simplification is the successful
computation of the spectrum of anomalous dimensions of the gauge
theory by mapping the relevant large $N$ matrix models to integrable
quantum spin chains \cite{min-zar, bs-long}. The matrix model in
question is nothing but the radial hamiltonian of the gauge theory
or the dilatation operator. The dilatation operator is, in general,
a complicated multi-matrix model whose Hamiltonian can only be
computed order by order in perturbation theory. However if one
focuses on protected operators of the gauge theory then the
dilatation operator takes on a particularly simple form: indeed it
is nothing but a sum of decoupled matrix harmonic oscillators
\cite{Beren-1,coll-1,beren-2}. Though the radial Hamiltonian (when
restricted to the sector of protected operators) appears to be a
non-interacting system, gauge fixing induces non-trivial
interactions among the microscopic degrees of freedom of the
Hamiltonian. Gauge fixing is necessary as the radial Hamiltonian
inherits a residual $U(N)$ gauge invariance from the original super
Yang-Mills theory. In the simplest case, when one studies the
dynamics of BPS operators built out of a single complex scalar even
the gauge fixed dynamics turns out to be free: indeed the gauge
fixed theory is nothing other than that of a collection of free
fermions \cite{llm, mandal-1, Beren-1,coll-1,beren-2,mandal-2}.
However this is not the generic scenario. Investigation of the
microscopic dynamics of  operators that involve several SYM fields
leads, in general, to interacting but integrable particle mechanics
that can be understood as generalizations of the celebrated Calogero
models.

In this present work, we study  this  connection between protected
operators in $\cal{N}$=4 SYM and Calogero models. The underlying
motivation is to develop the quantum many-body theories that are
relevant for the appropriate generalizations of the free fermion
picture of the Hamiltonian description of chiral primaries of the
gauge theory formed out of a single complex scalar. In particular,
we shall focus on the sub-sector of gauge theory formed out of three
complex scalar fields and two fermions known as the $su(2|3)$ sector
of $\cal{N}$ =4 SYM \cite{dyn}. The operator content of this
subsector is \beq\mathcal{W}_\alpha = \{Z_1,
Z_2,Z_3,\Psi_1,\Psi_2\}.\eeq $Z_1, Z_2, Z_3$ being three complex
chiral scalars and $\Psi_1, \Psi _2$ being two Fermions. The
motivation behind the choice of the  $su(2|3)$ sector is its closure
under dilatation as all local gauge invariant composite operators
formed out of these five fields \beq \mathcal{O} =
\prod_mTr(\mathcal{W}_{\alpha_1}\cdots \mathcal{W}_{\alpha_m})\eeq
mix only with each other to all orders in perturbation theory
\cite{dyn}. Before elaborating further on the technical details of
the dynamics of the protected operators contained in the $su(2|3)$
sector, it is worth laying out a summary of the basic results
obtained in the paper. We construct a gauge fixed version of the
tree level dilatation operator, which is nothing but a sum of matrix
harmonic oscillators and realize it as a generalization of the
celebrated supersymmetric Calogero model known as the Euler Calogero
model. Not all the excitations of the Euler Calogero model
correspond to BPS excitations of the the gauge theory, however, the
manifest supersymmetry of the Euler Calogero model can be utilized
to pick out those excitations which do correspond to the protected
gauge theory operators. This construction has been carried out in
chapter-5 of the paper. The $su(2|3)$ sector also contains a smaller
closed subsector of operator mixing namely the so called $su(1|1)$
sector containing a  scalar and a single Fermionic field. Within
this subsector, we have been able to construct a set of states of
the tree level dilatation operator that are protected in the large
$N$ limit, but have small anomalous dimensions at finite values of
the rank of the gauge group. The gauge theory operators
corresponding to these states interpolate between the so called
LLM{\cite{llm}} and BMN{\cite{bmn}}operators and provide us with a
set of non-BPS operators about which one can make
non-perturbative/all loops statements by studying the large $N$
limit of the corresponding tree level dilatation operator. For these
operators, we are able to recast the gauge fixed large $N$
dilatation operator as the well known rational supersymmetric
Calogero model. The construction of the Calogero model and the use
of its integrability to completely solve for its dynamics, enumerate
the degeneracies of the corresponding gauge theory operators and to
study the Yangian symmetry underlying their dynamics has been done
in chapters 2,3 and 4.

To recapitulate the free fermion picture of half BPS states it is
worth recalling that one can pick any one of the complex scalars,
say $Z_1$ of the $su(2|3)$ sector of the gauge theory and one then
has the standard result that all the operators formed out of only
$Z_1$'s are half BPS or chiral primaries of the gauge theory i.e
they do not have anomalous dimensions. The dilatation operator in
the half BPS sector of chiral primaries is then the tree level
dilatation operator, or the matrix Harmonic oscillator \beq H_{cp}
=\tr (A^{1\dagger }A_1).\eeq It is understood that one has mapped
the chiral primaries to the states of the dilatation operator with
$A^{1\dagger}$ being the creation operator for the $Z_1$ type of
excitations. The matrix harmonic oscillator simply counts the number
of fields sitting inside the state. As suggested in \cite{Beren-1}
it is more sensible to think of the above Hamiltonian as a  gauged
matrix model: \beq H_{cp} = \frac{1}{2}\tr \left( (D_tX_1)^2 +
X_1^2\right): D_t X = \dot{X} + [G,X]\eeq where $G$ is a gauge
connection. The gauge invariance is simply inherited from original
gauge theory for which the dilatation operator is the radial
Hamiltonian. Fixing the gauge $G =0$ is a projection of the dynamics
on to the singlet states, and the Hamiltonian simply reduces to that
of a matrix oscillator. Following standard techniques the gauge
choice reduces the number of degrees of freedom of the gauge theory
from $N^2$ to $N$, and the gauge invariant microscopic dynamics of
the half BPS sector can be reformulated as the dynamics of the $N$
eigenvalues of the matrix $X_1$. The change of variables form the
matrix elements to the eigenvalues introduces a Jacobian, which can
be absorbed in a redefinition of the wave function which
subsequently becomes antisymmetric enabling us to interpret the
eigenvalues $x_i$ as fermions. The hamiltonian for the eigenvalues
is simply the free one \beq H_{CP} =
\frac{1}{2}\sum_i\left(-\frac{\partial ^2}{\partial x_i^2} +
x_i^2\right)\label{ff}\eeq One thus has an interpretation of the
gauge invariant degrees of freedom of the scalar half BPS sector of
$\cal{N}$=4SYM in terms of $N$ free fermions described by
(\ref{ff}). The free fermion picture has proved to be extremely
useful in understanding non-perturbative aspects of the AdS/CFT
correspondence. For instance a precise map between half BPS
geometries and the phase space density of free fermions has been
proposed by \cite{llm} while several large excitations of the free
Fermions have been related to BPS branes and giant gravitons in the
dual string theory \cite{bal-1,corl, Beren-1}.

The existence of two equivalent descriptions of the matrix harmonic
oscillator has also been viewed as an example of an exact
realization of open/closed duality in  $\cal{N}$ =4 SYM
\cite{Beren-1}. The description of the states of the matrix model in
terms of products of traces of the matrix creation operators has
been viewed  as a closed string description of the dual string
theory. An operator such as \beq (\mathcal{B}_n)^\dagger = \tr
((A^\dagger )^n)\eeq can be viewed as a creation operator for a
closed string mode of energy $n$. A typical matrix model state of
energy $n$ can then be described by all the partitions of the number
$n$ into $n_1 \cdots n_i$ such that $n_1 \geq n_2 \cdots \geq n_i$.
To each partition one may associate a Young Tableaux having columns
with $n_1 \cdots n_i$ boxes. This has been regarded as a realization
of the description of the degeneracy of the dual closed string
excitations on \cite{corl,Beren-1}. As a matter of fact a world
sheet/string sigma model  description of the matrix harmonic
oscillator has also been found recently \cite{ho-string}, though the
connection of the world sheet description of the matrix oscillator
and string theory on $AdS_5\times S^5$ probably requires further
study.

On the other hand, the description of the states of the matrix model
in terms of eigenvalues: the free Fermion picture has been regarded
as a an open string /D brane description of the half BPS sector of
the AdS/CFT correspondence. The classic Bosonization result that
relates the degeneracies of states of a free fermion system in $1+1$
dimensions to that of a chiral Boson, where once again the
degeneracies are counted by the number of partitions of the integer
energy levels has been regarded as  an open string description of
the BPS spectrum of the dual string theory.

Of special interest in recent investigations has been the
description of large excitations i.e excitations with energies of
$O(N)$. These large departures from the Fermi sea, known as the
giant gravitons correspond to operators built out of determinants
and sub-determinants rather than traces \cite{corl, Beren-1}. For
instance the gauge theory description of the maximal BPS giant
corresponds to the state \beq \epsilon _{i_i \cdots i_N}\epsilon
^{j_1 \cdots j_N} A^{1\dagger i_1}_{j_1} \cdots A^{1\dagger
i_N}_{j_N}|0>\eeq

Such large non-perturbative BPS excitations provide one with a gauge
theory description of BPS branes on the dual string geometry. The
free Fermion picture gives a particularly simple and elegant
description of the BPS giants:  they simply correspond to exciting
an eigenvalue from the bottom of the Fermi sea to the top. The
Fermionic description sheds light on a host of issues related to the
AdS/CFT correspondence. For instance the vibration frequencies of
(BPS) giant gravitons  computed by a world-volume
computation\cite{das-1,das-2}can be reproduced in the gauge theory
language using by solving the matrix harmonic oscillator. The matrix
oscillator description can also be used to clarify a host of issues
regarding the gauge theory duals of non-perturbative string states.
Several such  recent interesting developments have been discussed in
\cite{saff-1, saff-2,saff-3,saff-4,saff-5,saff-6,saff-7,oku}.

The class of BPS operators described in the brief review above are
all charged under a $U(1)$ of the $SO(6)$ R symmetry group of the
gauge theory. However, it is just as natural to consider operators
that carry several other charges. These would correspond to
protected operators that involve more than a single scalar field
inside a trace. Hence, a natural question that arises out of this
line of investigation is how the free fermion picture changes in a
systematic way once BPS excitations involving multiple fields are
considered. For instance, in the particular case of the $su(2|3)$
sector considered here, one could have operators such as\beq
tr(Z_1^n\Phi )\eeq where $\Phi $ can be any one of $Z_2, Z_3,\Psi
_1, \Psi _2 $. These operators, being the supersymmetry descendants
of $\tr Z_1^n$ are also protected.  Maximal giant gravitons such as
\beq \epsilon_{i_1 \cdots i_N}\epsilon ^{j_1 \cdots j_N}(A^{\dagger
1})^{i_1}_{j_1}\cdots (A^{\dagger 1})^{i_{N-1}}_{j_{N-1}}(A^{\dagger
\alpha})^{i_{N}}_{j_{N}}|0>\eeq where $A^{\dagger \alpha}$
corresponds to an impurity excitation also fall into the same
category of protected operators. Similarly, one could build
protected operators with multiple 'impurity ' fields inside a single
trace. A particularly simple example would be \beq \tr ( Z_1\Phi
\Phi ).\eeq The principal question that we shall address in this
paper is what the appropriate generalization of the free Fermion
picture is when generic protected operators that involve an
arbitrary number of fields are considered. This question is also
relevant from the point of view of understanding the role of
supersymmetry in the description of BPS dynamics as many-body
theories. Since the $su(2|3)$ sector contains some amount of
supersymmetry one can hope to learn how the supersymmetry manifests
itself in the open-string picture. The role of supersymmetry is not
obvious at the level of the free Fermion system\footnote{For a
recent parallel line of investigation into the study of multi-charge
giants,
see \cite{yon-mul-ch}}. 

The  dilatation operator, restricted to the set of BPS operators in
the $su(2|3)$ sector is nothing but the sum of five decoupled
harmonic oscillator Hamiltonians. \beq H' = \sum_{i =1 }^3
\tr\left( A^{i \dagger }A_i \right) + \frac{3}{2}\sum_{I =1 }^2
\tr\left( \Psi^{I \dagger }\Psi_I \right).\eeq The factor of
$\frac{3}{2}$ in front of the Fermionic Hamiltonian is nothing but
the engineering dimension of the Fermionic fields of the gauge
theory. In what is to follow, we shall subtract a term proportional
to the Fermion number operator and work with\beq H = H' -
\frac{1}{2}\sum_{I =1 }^2 \tr\left( \Psi^{I \dagger }\Psi_I
\right).\eeq Since the dilatation operator does not change the
Fermion number, $H$ and $H'$ carry the same information, and in
various analyses that are carried out in this paper, we shall give
explicit prescriptions for understanding various features (such as
degeneracies, Yangian symmetries etc)  of $H'$ from the studies of
the corresponding properties of $H$. To simplify the notation we
shall write the Hamiltonian as\beq H = \sum_{\alpha =1 }^5\tr\left(
A^{\alpha \dagger }A_\alpha \right).\eeq It will be understood that
$\alpha = 1,2,3$ correspond to Bosonic matrices while $\alpha = 4,5$
correspond to Fermionic ones.

Generalization of the closed string point of view from the half BPS
sector involving a single matrix oscillator to the $su(2|3)$ sector
follows immediately. The closed string excitations are identified as
states of the matrix model formed out of products of traces of the
creation operator acting on the vacuum, i.e., they are states of the
form\beq \prod_n\tr \left(A^{\dagger \alpha _1} \cdots A^{\dagger
\alpha _n}\right)|0>\eeq Of course when all the $\alpha _i =1 $, we
revert back to the half BPS sector of chiral primaries. In this
point of view one can work out the degeneracies corresponding to a
states with given energies much along the same lines as the analysis
involving a single field $Z_1$. The analysis is a little more
involved but it can be carried out nevertheless using the Polya
formulae for counting the number of distinct 'words' formed out of a
certain number of 'letters' in an 'alphabet': in our case
five\footnote{We shall refer the reader to \cite{sem-polya} for a
simple discussion of Polya counting applied to matrix harmonic
oscillators}. However the open string description i.e the analog of
the eigenvalue dynamics of the multi-matrix model is not obvious at
all. An open string description of the full $su(2|3)$ sector would
require an understanding of how the free Fermion picture of the
chiral primary states changes in a systematic way once multi-charge
BPS excitations are allowed. Such BPS excitations would correspond
to the impurity fields $\alpha = 2\cdots 5$ to be present inside a
single trace, the simplest of which would correspond to states such
as\beq\tr\left ((A^{1\dagger })^n A^{\alpha \dagger}\right)|0> \eeq
where $\alpha $ is one of the impurity fields $2\cdots 5$. As
mentioned before, one can in general have BPS states with a large
number of impurity fields inside a single trace. Clearly,  the
problem of deriving an open string description of the multi-charge
BPS states amounts to finding a description of the dynamics of the
eigenvalues of the matrix $X_1$ in the background of the impurity
fields.

In the present work we shall take a step in the direction of
understanding this problem. We shall be able to show that in the
presence of impurity excitations the eigenvalues can be understood
as Fermions with internal/spin degrees of freedom. They will turn
out to  interact with each other through spin dependent inverse
square interactions. As a matter of fact we shall be able to
formulate the dynamics of the eigenvalues in the background of a
arbitrary number of impurity excitations in terms of generalizations
of the celebrated Calogero systems. Furthermore, we shall also be
able to show that for case of the simplest departure from the Free
Fermion picture involving the study of BPS operators consisting of a
single impurity field inside a trace the dynamics reduces to the
well known super symmetric rational Calogero model. We shall study
this sector in some detail, as it has all the features of the most
general particle mechanics that one can encounter in the study of
the multi-charge BPS operators. The relation between super-Calogero
models and matrix models was made by Dabholkar \cite{dab-mar}, where
the super Calogero model was shown to be a consistent truncation of
the Marinari-Parisi model. In the case at hand we shall be able to
see that a similar truncation has a natural interpretation in the
study of $\mathcal{N} =4$ SYM as the restriction of the dynamics of
the dilatation operator to protected operators of a particular type.
After setting up the correspondence between multi charge operators
and the super Calogero system we shall recover the complete spectrum
and degeneracies of the BPS excitations of the matrix model within
the framework of the Calogero system. This might be regarded as  the
realization of an open/closed duality for multi-charge protected
operators much along the same lines as the  one between the  single
matrix oscillator and the free Fermion system.

Not all the excitations of the matrix model correspond to protected
operators of the gauge theory. However, the non-BPS excitations of
the matrix model are bona-fide local composite operators of the
gauge theory. Other than the sector of protected operators, the
matrix oscillators also provide one with a non-perturbative
definition of the gauge theory dilatation operator in the limit of
$g^2_{YM}\rightarrow 0$. As is well known, the string dual to the
free gauge theory is notoriously hard to pin down. Thus the ulterior
motive behind our study of the tree level dilatation operator is
that perhaps its gauge fixed form can be utilized to discover the
string theory which is relevant in the limit of zero Yang-Mills
coupling. Though we do not make an attempt at finding the string
theory, we do identify and study operators that have the curious
property of  being protected in the large $N$ limit, while at finite
values of $N$ they turn out to be BMN like operators with small
anomalous dimensions. The parameter that governs their BPS condition
is $\frac{1}{N}$. It is in the study of these operators that we find
that the dilatation operator takes on the familiar form of the
Calogero model.

Apart from analyzing the spectrum and the open/closed duality, we
shall also use the Calogero system to investigate the hidden
symmetries that lead to its integrability. The motivation for doing
this is the use of the protected sector of the gauge theory as a
probe to understand whether or not any of the integrable structures
(such as Yangian symmetries) that are present in the string sigma
model survive the supergravity limit. Interestingly enough, for the
case of the Calogero model we shall be able to see that the
underlying symmetry is not an Yangian but rather its loop algebra.
In the light of the fact that the loop algebra can be regarded as a
classical limit of the Yangian algebra (the symmetry of the string
sigma model) it is reasonable to expect it to be the symmetry of the
classical limit of the string theory. In the simplest non-trivial
example that we study in this paper, this expectation is indeed
realized.

After a detailed description of  operators involving a single
impurity field inside a matrix trace in terms of the Calogero model,
we shall describe the dynamics of the most general (multi-charge)
protected operators. The particle mechanics in the general case will
turn out to be governed by a particlar (integrable) generalization
of the rational Calogero systems known as the Euler-Calogero
systems \cite{fer-mar}. We shall be able to exploit the integrability
of these systems to understand the spectrum and degeneracies of the
most general multi-charge BPS operators as well.

We shall finally conclude with  comments on some unresolved issues
and directions for future explorations. 

\section{Multi-Matrix Harmonic Oscillators:}
In this section we shall present an overview of the techniques that
are necessary to have a gauge fixed description of a collection of
matrix harmonic oscillators. The starting point is  a system of $d$
Hermitian matrices, $(X^\alpha)^i_{j}$, $\alpha = 1\cdots d$. $i,j
=1 \cdots N$. Keeping in mind the $su(2|3)$ sector, we shall let
$d=5$, with $d=1\cdots 3$ being bosonic and the rest fermionic. The
Hamiltonian for the matrix model will be taken to be a sum of
harmonic oscillators, \beq H = \sum_\alpha \tr\frac{1}{2}\left( \Pi
^\alpha \Pi ^\alpha + X^\alpha X^\alpha \right)\eeq$\Pi $ is the
momentum conjugate to $X$, and the canonical commutation relations
are, \beq [(X^\alpha)^i_{j} , (\Pi ^\beta) ^k _l ]_\pm =i\hbar
\delta ^{\alpha , \beta } \delta ^k_{j}\delta ^i_{l}\eeq One could
go to the Holomorphic basis of creation and annihilation operators,
in which the Hamiltonian becomes \beq H =\sum_\alpha
tr\left(A^{\dagger \alpha}A_\alpha \right)\eeq

\subsection{Generalized Calogero Systems:} We want to write the
system of Harmonic oscillators in a basis in which one of the
matrices, $X^1$ is diagonal. Changing variables from the matrix
elements to the eigenvalues in matrix models involving several
matrices is in general hard to accomplish. However, when only one
matrix is diagonalized this becomes tractable. The matrices do not
couple to each other, so the dynamics of the eigenvalues of a single
matrix is that of a spin-Calogero type, where the role of spin is
played by the generators of unitary conjugations \cite{min-cal-1}.
This creates an effective coupling to the remaining matrices, as the
Gauss law relates these generators to those of the remaining
matrices. Such a reduction was worked out by Ferretti in
\cite{fer-mar} in the context of the Marinari-Parisi model(see
also\cite{fer-raj}). Below we outline the procedure for our case.

Let us denote the diagonal elements of $X^1$ by $x_i$
\beq X^1 = U^\dagger x U.\eeq
Furthermore, let us denote the oscillators in
this basis by lower case letters \beq (a^\alpha )^i_j = (UA^\alpha
U^\dagger )^i_j,(a^{\dagger \alpha} )^i_j = (UA^{\dagger \alpha
}U^\dagger )^i_j. \alpha \neq 1\label{newbasis}\eeq
Let us now
proceed to write down the Hamiltonian of the decoupled set of
oscillators as a generalized Calogero system. We are going to treat
all the oscillators other than the first one as impurities so
\beq H
= H_1 + H_{Imp}\eeq where $H_1$ denotes the Hamiltonian for the
first oscillator. $H_{Imp}$ can be written easily enough as \beq
H_{Imp} = \sum_{\alpha \neq 1} \tr(a^{\dagger \alpha}a_{\alpha
}).\eeq To write the first oscillator in the eigenvalue basis one
starts with the metric on the space of Hermitian matrices \beq ds^2
= \sum_i dx_idx_i + \sum _{i\neq j} (x_i - x_j)^2{\omega }^{\star
p}_q \omega ^{ q}_p.\eeq The one forms $\omega $ are defined as\beq
\omega ^i_j = (dU)^i_k(U^\dagger )^k_j.\eeq Similarly, one also has
the dual vector fields $\mathcal{L}$ \beq \mathcal{L}^i_j =
U^i_m\frac{\partial }{\partial U^m_j}\eeq that obey the $U(N)$ Lie
algebra\beq[\mathcal{L}^i_j, \mathcal{L}^k_l] = \delta ^k_j
\mathcal{L}^i_l -\delta ^i_l \mathcal{L}^k_j\eeq Using the metric
the momentum operator can be written as
\beq \frac{
\partial}{\partial X^j_i} = (U^\dagger )^i_k\pi ^k_lU^k_j\eeq
where\beq\pi ^i_j = \frac{\partial}{\partial x_i}\delta ^i_j +
\frac{1-\delta ^i_j}{x_i - x_j}\mathcal{L}^i_j \eeq
We can now write
\beq H_1 = \sum_i\frac{1}{2}\left(-\frac{\partial }{\partial
x_i^2} + x_i^2\right) + \frac{1}{2} \sum_{i\neq j}\left(
\frac{\mathcal{L}^i_j\mathcal{L}^j_i}{(x_i - x_j)^2}\right)\eeq
This is clearly a generalized $U(N)$ spin-Calogero system. However, we
want to formulate the particle mechanics completely in terms of the
microscopic degrees of freedom which the are $N$ eigenvalues $x_i$
and the remaining matrix oscillators $a^{\alpha i}_j, a^{\dagger
\alpha i}_j, \alpha \neq 1$. To do that we note that $U(N)$ singlet
states of the particle mechanical system would generically be of the
kind\beq \Psi^{i_1 \cdots i_n}_{j_i \cdots j_n}(x)\Pi _{k= 1}^n
(a^{\dagger \alpha _k}) ^{j_k}_{i_k} |0>.\eeq $\Psi $ is an $U(N)$
tensor which depends on the $N$ eigenvalues $x_i$. The dependence of
the state on the $\frac{N(N-1)}{2}$ angular degrees of freedom is
contained in $(a^{\dagger \alpha })^i_j$ which depend on the angular
coordinates through (\ref{newbasis}). It may now be easily verified
that \beq[\mathcal{L}^i_j,(a^{\dagger \alpha })^a_b] = [\sum
_\beta\left( (a^{\dagger \beta })^i_l(a^{\beta })^l_j - (a^{\dagger
\beta })^l_j(a^{\beta })^i_l\right),(a^{\dagger \alpha })^a_b].\eeq
This identity follows from noticing that\beq
[\mathcal{L}^i_j,(a^{\dagger \alpha })^a_b] = \delta ^a_j
(a^{\dagger \alpha })^i_b -\delta ^i_b (a^{\dagger \alpha })^a_j\eeq
which may be compared with the explicit action of the angular
derivatives on the angular coordinates present in the definition of
$(a^{\dagger \alpha })^a_b$. We may thus replace the vector fields
appearing in the Hamiltonian by the matrix operators, i.e.
\beq
\mathcal{L}^i_j =
 \sum _\beta\left( (a^{\dagger \beta })^i_l(a^{\beta })^l_j -
(a^{\dagger \beta })^l_j(a^{\beta })^i_l\right)\label{vecosc}\eeq
From now on it will always be implied (unless stated explicitly)
that the vector fields have been replaced by their oscillator
realization (\ref{vecosc}). We have thus completed writing the
Hamiltonian in terms of the degrees of freedom available to us in
the basis in which the first matrix is diagonal. The Hamiltonian
being
\beq H =\sum_i\frac{1}{2}\left(-\frac{\partial }{\partial
x_i^2} + x_i^2\right) + \frac{1}{2} \sum_{i\neq j}\left(
\frac{\mathcal{L}^i_j\mathcal{L}^j_i}{(x_i - x_j)^2}\right) +
\sum_{\alpha \neq 1} \tr(a^{\dagger \alpha}a_{\alpha })\eeq with \beq
\mathcal{L}^i_j =
 \sum _{\beta \neq 1}\left( (a^{\dagger \beta })^i_l(a^{\beta })^l_j -
(a^{\dagger \beta })^l_j(a^{\beta })^i_l\right)
\label{euler-calogero}\eeq

\subsection{Residual Constraints on Physical States:}
A typical state $| \psi >$ of the
Calogero system is
\beq | \psi > = \psi ^{i_i \cdots i_m}_{j_i\cdots
j_m}(x)(a^{\dagger \alpha _1})^{j_1}_{i_1}\cdots (a^{\dagger \alpha
_1})^{j_m}_{i_m}|0>,\eeq
where $\psi $ is a $U(N)$ tensor. Not all
the states of the many-body theory are allowed states of the gauge
fixed matrix model. The states have to be invariant under the
residual gauge symmetry left over even after carrying out the $U(N)$
rotation to the space of eigenvalues of $X^1$.  One must ensure that
the diagonal subgroup of $U(N) = U(1)^N$ that leaves the eigenvalues
invariant also leave the state invariant. So physical states have to
satisfy the constraint
\beq \mathcal{L}^i_i | \psi >  = \sum _\beta\left(
(a^{\dagger \beta })^i_l(a^{\beta })^l_i - (a^{\dagger \beta
})^l_i(a^{\beta })^i_l\right) | \psi > = 0\eeq

The model described above can be regarded as a generalization of the
well known spin-Calogero models.  Unlike the usual Calogero models
the model above has a very large number of `spin' degrees of
freedom.  The model is still Fermionic, as the overall wave function
is antisymmetric under the exchange of the particles. Thus the Free
fermion picture of BPS operators carrying a single $U(1)$ charge
seems to be a replaced by a picture of interacting Fermions. The
Fermions carry an internal spin degree of freedom and interact
through spin dependent inverse square interactions.

The classical limits of such generalized $SU(N)$ Calogero systems
have been studied in the literature in the past for independent
reasons and they are referred to as Euler- Calogero systems. We
shall adhere to this terminology in the present work as well. These
systems are also known to be integrable at the classical
level \cite{eul-cal-1, eul-cal-2}. Later in paper, we shall be able
to utilize the connection to matrix oscillators to confirm the
quantum integrability of these models and understand their spectrum.

The $SU(N)$ Calogero model is known to contain various Calogero
models with fewer number of spin degrees of freedom as  consistent
truncation of its dynamics to suitable chosen subspaces of its full
Hilbert space. For a discussion of such reductions in the context of
trigonometric Calogero models we shall refer to \cite{min-cal-1,
min-cal-2, alex-review}. Thus it is of interest to study whether or
not the usual Calogero models play any special role in the
understanding of BPS operators of the gauge theory. In the following
section we shall show that this is indeed true.
\section{A Dabholkar-like Truncation:} The first
class of operators that we shall look at are the ones that have at
the most only a single impurity excitation located inside a single
trace. Moreover, we shall restrict ourselves to the case where the
impurities are Fermionic. These are states of the form \beqs
\frac{1}{\sqrt{N^m}}\tr \left((A^{\dagger 1})^m \right)
\frac{1}{\sqrt{N^{n_1+1}}}\tr\left((A^{\dagger
1})^{n_1}\Psi^{\dagger \alpha _1}\right)  \cdots
\frac{1}{\sqrt{N^{n_i+1}}}\tr\left((A^{\dagger 1})^{n_i}
\Psi^{\dagger \alpha _i}\right)|0>. \label{Simp}\eeqs $\alpha _1
\cdots \alpha _i = 1,2$. An interesting aspect of these states is
that they are protected in the large $N$ limit.

These states when written in the basis in
which $X^1$ is diagonal would generically appear as
\beq\prod_m\Psi(x_1 \cdots x_N)^{i_1 \cdots
i_m}(\mathcal{A}^{\dagger \alpha _1})_{i_1}\cdots
(\mathcal{A}^{\dagger \alpha _m})_{i_m}|0> + O(\frac{1}{N})
\label{Dhab}\eeq
where, \beq (\mathcal{A}^{\dagger \alpha })_i = (\psi^{\dagger \alpha
})^i_i.
\eeq
are the excitations corresponding to the diagonal
matrix elements of the impurity creation operators in the rotated
basis. It is possible to perform a consistent truncation of the
particle mechanical system to a Hilbert space $\mathcal{H}_d$
spanned by states of the above type.

To see that this truncation is consistent one needs to show that
$\mathcal{H}_d$ is closed under the action of the Hamiltonian. This
is obviously the case as the Hamiltonian does not change the number
of impurity fields inside the traces.

As the $su(2|3)$ sector has two Fermionic degrees of freedom, one
can consider states for which the impurities correspond to only one of
the two Fermionic degrees of freedom available to us, that is,
either $\Psi^1$ or $\Psi^2$, which we will simply call $\Psi$.
This is the so-called $su(1|1)$ sector of the
gauge theory, and operators formed out of the two degrees of freedom,
$X^1$ and $\Psi$, are also closed under dilatation.
Furthermore, the quartic spin interaction term of the Euler-Calogero
model assumes a much simpler and familiar form within this truncated
subspace, as it can be represented by a graded exchange operator
\beq \mathcal{L}^i_j\mathcal{L}^j_i= \frac{1}{2}(1 - \Pi_{i,j}).\eeq
$\Pi_{i,j}$ is a graded permutation operator that exchanges the
spins at the lattice sites $i$ and $j$ while picking up a negative
sign if both the spins happen to be Fermionic.

To see how this arises, assume that the angular $SU(N)$ conjugation
generators $\mathcal{L}$ are in a representation generated by
\beq
\mathcal{L}_j^i = b_i^\dagger b_j - f_j^\dagger f_i
\eeq
where $b_i , b_i^\dagger$ and $f_i , f_i^\dagger$ are a set of bosonic
and fermionic oscillator ladder operators, respectively. The above
construction embeds in the oscillators' Fock space all totally symmetric
products of the fundamental times all totally antisymmetric products
of the antifundamental of $SU(N)$. The residual physical constraint reads
\beq
\mathcal{L}_i^i | \psi > = (b_i^\dagger b_i - f_i^\dagger f_i )
| \psi > = 0
\eeq
which implies that the Boson and Fermion number for each index $i$
are both equal to $0$ or $1$. This realizes the group $SU(1|1)$ on each
site $i$, acting upon the `spin' states of the site labelled by their
Fermion number $0,1$. Using the above condition, it is an easy
matter to show that $\mathcal{L}_j^i \mathcal{L}_i^j$ reduces to
the graded exchange operator $1-\Pi_{i,j}$ when it acts on physical
states.

To complete the demonstration, we remark that the representation
of $\mathcal{L}_j^i$ carried by the states (\ref{Simp}) is exactly
the one embedded in the above construction. Indeed, writing
$A^{\dagger 1} = A^\dagger$, gauge invariant
states in terms of $b_i^\dagger$ and $f_i^\dagger$ are generated
through the action of operators
\beq
b_i^\dagger (A^\dagger)^n_{ij} f_j^\dagger = \tr \left(
(A^\dagger)^n f^\dagger b^\dagger \right)
\eeq
where we view $b_i^\dagger$ as a row vector and $f_i^\dagger$
as a column vector. This is identical to the operators appearing
in (\ref{Simp}) upon identifying $\Psi^\dagger$ with $f^\dagger
b^\dagger$ (both operators are fermionic and have the same
$SU(N)$ transformation properties). In this realization, however,
there are no multiple impurities per trace, since
\beq
\tr \left( (A^\dagger)^n f^\dagger b^\dagger (A^\dagger)^m
f^\dagger b^\dagger \right) = \tr \left( (A^\dagger)^n f^\dagger
b^\dagger \right) \tr \left( (A^\dagger)^m
f^\dagger b^\dagger \right)
\eeq
So the space spanned by single impurity traces is isomorphic to
the above $SU(1|1)$ spin representation. Further, in the $X^1$
diagonal (eigenvalue) representation, physical states arise
through the action of $b_i^\dagger f_i^\dagger$ for each
eigenvalue. We can thus identify $(\mathcal{A}^\dagger )_i
= (\Psi^\dagger )_i^i$
with the above operator, obtaining a correspondence with
Dhabolkar-like states (\ref{Dhab}).

By using the formalism developed above, the Hamiltonian in the
$SU(1|1)$ sector can be written as \beq H =
\sum_i\frac{1}{2}\left(-\frac{\partial }{\partial x_i^2} +
x_i^2\right) + \frac{1}{2} \sum_{i\neq j}\left( \frac{1 -
\Pi_{i,j}}{(x_i - x_j)^2}\right) + \sum_j \mathcal{A}^{\dagger j}
\mathcal{A}_j \eeq By using the fermionic form of the graded
permutation operator, \beq \frac{1}{2}\left(\mathcal{A}^{\dagger
i}\mathcal{A}_i + \mathcal{A}^{\dagger j}\mathcal{A}_j-
\mathcal{A}^{\dagger i}\mathcal{A}_j - \mathcal{A}^{\dagger
j}\mathcal{A}_i\right) = 1-\Pi_{i,j}\eeq we can recast the above
Hamiltonian as \beq H = \sum_i\left(-\frac{1}{2}\frac{\partial
}{\partial x_i^2} + \mathcal{A}^{\dagger i}\mathcal{A}_i +
\frac{1}{2}x_i^2\right) + \frac{1}{2} \sum_{i\neq j} \left(
\frac{\mathcal{A}^{\dagger i}\mathcal{A}_i - \mathcal{A}^{\dagger
i}\mathcal{A}_j}{(x_i - x_j)^2}\right) \eeq This is nothing but the
supersymmetric rational Calogero model. This very model has appeared
in the analysis of superstrings in two dimensions where it
was shown by Dabholkar to be a consistent truncation of the
Marinari-Parisi model \cite{dab-mar}. The truncation that we perform
is similar to the one carried out by Dabholkar, however, it is to be
kept in mind that the eigenstates of the Calogero system correspond
to protected operators of the gauge theory only in the large $N$
limit. Another gratifying aspect of the present analysis, which will
be made clear in the following sub-section is that one can have a
one to one map between the excitations of the Calogero system and
the those of the matrix model. Such a map between the open and
closed string pictures is slightly obscure in the approach pioneered
in \cite{dab-mar}.

We thus see that the super-Calogero
model is relevant to the study of $\cal{N}$=4 SYM as being the
natural generalization of the theory of free Fermions which is
relevant for the study of BPS operators with no impurities. The
Calogero model is still a theory of Fermions as the overall wave
function is antisymmetric under the exchange of the particles, but
the Fermions are no longer free and they carry an internal spin
degree of freedom.

\subsection{From the Matrix Model to the Calogero System:}
We shall now elaborate on the connection of the Calogero model to
the matrix model, and in the process provide an alternative
explanation for why it was reasonable to replace the quartic spin
interaction term by the quadratic graded permutation operator. The
simplest way to understand the connection to the super Calogero
model is by looking at the spectrum of the matrix model in the
subspace considered above. It is quite clear that the spectrum of
the matrix model is the same as that of a system of Bosonic and
Fermionic oscillators with frequencies given by integers. One can
introduce Bosonic an Fermionic creation operators $B_n$ and $F_k$
which create oscillator states of energies $n $ and $k$
respectively.  It is then possible to map the matrix model states to
oscillator states using the following identification: \beq B_n |0>
\leftrightarrow \frac{1}{\sqrt{N^n}}tr(A^{\dagger })^n |0>\eeq for
Bosonic states and \beq F_k |0> \leftrightarrow
\frac{1}{\sqrt{N^k}}\left[tr(A^{\dagger })^{k-1}\Psi ^\dagger
\right]|0>\eeq for the Fermionic ones.  The Bosonic and Fermionic
oscillators can be taken to be related to each other through a
supersymmetry algebra given by: \beqs & &[F_m,F_n]_+ = 0, [B_m,F_n]=
0,
[B_m,B_n]=0\nonumber \\
& &[Q, F_m]_+ =0, [Q^\dagger, F_n]_+ = B_n, [H,F_n] = nF_n\nonumber
\\
& &[Q, B_n] = 2nF_n, [Q^\dagger, B_n] =0, [H,B_n] =nB_n
\label{super-al}\eeqs $H$ in the above set of equations is the
Hamiltonian for the free super oscillators whose frequencies are
given by integers. But this is nothing but the rational
super-Calogero model in disguise. The super-Calogero model and its
spectrum has been studied in various papers in the past, see for
example \cite{ghosh-cal, freedman-cal}, and it is known that it can
be brought to a form where the Hamiltonian becomes a collection of
free super oscillators by  a similarity transform.  We shall now
summarize the similarity transformation that brings the Calogero
model to the form of the super-oscillators  for the sake of
completeness.

The Calogero model has a manifest supersymmetry which is generated
by\beqs \mathcal{Q} = \sum_i \mathcal{A} ^{\dagger i} \Pi_i\nonumber
\\\mathcal{Q}^\dagger  = \sum_i \mathcal{A}_i \Pi ^\dagger _i\eeqs
where $\Pi _i $ are the coupled momentum
operators \cite{exchange} \beq \Pi _i =  p_i -iW_i, \Pi ^\dagger _i =
p_i +iW_i.\eeq $p_i = -i\frac{\partial }{\partial x_i}$ while $W_i =
\frac{\partial W}{\partial x_i}$ with $W$ being the superpotential
\beq W = -\ln\Pi_{i<j}(x_i - x_j) + \frac{1}{2}\sum_i x_i^2\eeq Some
straightforward algebra shows that (up to a constant term) the
Hamiltonian can be written in a manifestly supersymmetric form\beq H
= \frac{1}{2}[\mathcal{Q}, \mathcal{Q} ^\dagger ]_+\eeq

The ground state has Fermion number $= 0$, and it is the same as
that of the free Fermion system:\beq \Omega =
\prod_{i<j}(x_i-x_j)e^{-\frac{1}{2}\sum_ix_i^2}|0>\eeq

The higher excitations above this ground state can be understood in
a purely algebraic fashion by mapping the Calogero system to a
system of free super-oscillators with frequencies given by integers
$1\cdots N$. The explicit form of the similarity transformation that
maps the super-Calogero system to the system of free
super-oscillators has been worked out in detail in \cite{ghosh-cal},
and we shall gather together the relevant results that are necessary
for understanding the degeneracies. One can introduce the Bosonic
and Fermionic raising operators$B_n$ and $F_n$\beq \frac{1}{2^n}B_n
=\sum_i \Gamma ^{-1}x^n_i\Gamma, \frac{1}{2^{n-1}}F_n = \sum_i
\Gamma ^{-1}\mathcal{A}^\dagger _ix^{n-1}_i \Gamma,\eeq where\beq
\Gamma = e^{\frac{S}{2}}(-\ln \Omega)\eeq and \beq S =
\frac{1}{2}\sum_i \frac{\partial ^2}{\partial x_i^2} + \sum_{i\neq
j}\frac{1}{x_i-x_j}\frac{\partial }{\partial x_i} - \sum_{i\neq
j}\frac{1}{(x_i-x_j)^2}(\mathcal{A}^\dagger _i\mathcal{A} _i
-\mathcal{A}^\dagger _i\mathcal{A} _j)\eeq Similarly, one could also
apply the similarity transformation to the supercharges,\beq Q =
\Gamma ^{-1} \mathcal{Q} \Gamma, Q^\dagger = \Gamma ^{-1}
\mathcal{Q}^\dagger \Gamma\eeq Some straightforward but lengthy
algebraic computations yield that the algebra obeyed by the raising
operators and the supercharges is (\ref{super-al}). Thus, we see
that the truncation of the matrix model to states involving only one
impurity inside a single trace can be described by the
super-Calogero model.
\subsection{Degeneracies} This algebraic structure makes the spectrum and the associated degeneracies
of the model extremely transparent. As in the free Fermion picture
the degeneracies of the Bosonic states are counted by partitions of
integers. The states \beq B_n|0> \mbox{and} \prod_{i=l}^l
B_{n_i}|0>, \sum_i n_i =n\eeq are degenerate which is the open
string description of the degeneracies between matrix model
states\beq \tr[(A^\dagger)^n]|0> \mbox{and } \prod
_i\left[\tr(A^\dagger)^{n_i}\right]|0>, \sum_i n_i = n.\eeq Making a
choice of ordering such that $n_1 \geq n_2 \geq n_3 \cdots $ one has
the result that the states with energy $n$ can be represented by
Young diagrams with $n$ boxes. For instance the state with energy
$n$ corresponds to a Young diagram with columns of length $n_1 \geq
n_2 \geq n_3 \cdots $.
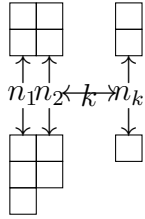
\begin{center}
\begin{pgfpicture}{-1cm}{-0.5cm}{3cm}{3cm}
\pgfrect[stroke]{\pgfpoint{0cm}{0cm}}{\pgfpoint{10pt}{10pt}}
\pgfrect[stroke]{\pgfpoint{0cm}{10pt}}{\pgfpoint{10pt}{10pt}}
\pgfrect[stroke]{\pgfpoint{0cm}{20pt}}{\pgfpoint{10pt}{10pt}}
\pgfputat{\pgfpoint{5pt}{35pt}}{\pgfbox[center,center]{$\downarrow$}}
\pgfputat{\pgfpoint{5pt}{45pt}}{\pgfbox[center,center]{$n_1$}}
\pgfputat{\pgfpoint{5pt}{55pt}}{\pgfbox[center,center]{$\uparrow$}}
\pgfrect[stroke]{\pgfpoint{0cm}{60pt}}{\pgfpoint{10pt}{10pt}}
\pgfrect[stroke]{\pgfpoint{0cm}{70pt}}{\pgfpoint{10pt}{10pt}}
\pgfrect[stroke]{\pgfpoint{10pt}{10pt}}{\pgfpoint{10pt}{10pt}}
\pgfrect[stroke]{\pgfpoint{10pt}{20pt}}{\pgfpoint{10pt}{10pt}}
\pgfputat{\pgfpoint{15pt}{35pt}}{\pgfbox[center,center]{$\downarrow$}}
\pgfputat{\pgfpoint{15pt}{45pt}}{\pgfbox[center,center]{$n_2$}}
\pgfputat{\pgfpoint{15pt}{55pt}}{\pgfbox[center,center]{$\uparrow$}}
\pgfrect[stroke]{\pgfpoint{10pt}{60pt}}{\pgfpoint{10pt}{10pt}}
\pgfrect[stroke]{\pgfpoint{10pt}{70pt}}{\pgfpoint{10pt}{10pt}}
\pgfputat{\pgfpoint{25pt}{45pt}}{\pgfbox[center,center]{$\leftarrow$}}
\pgfputat{\pgfpoint{30pt}{45pt}}{\pgfbox[center,center]{$k$}}
\pgfputat{\pgfpoint{35pt}{45pt}}{\pgfbox[center,center]{$\rightarrow$}}
\pgfrect[stroke]{\pgfpoint{40pt}{20pt}}{\pgfpoint{10pt}{10pt}}
\pgfputat{\pgfpoint{45pt}{35pt}}{\pgfbox[center,center]{$\downarrow$}}
\pgfputat{\pgfpoint{45pt}{45pt}}{\pgfbox[center,center]{$n_k$}}
\pgfputat{\pgfpoint{45pt}{55pt}}{\pgfbox[center,center]{$\uparrow$}}
\pgfrect[stroke]{\pgfpoint{40pt}{60pt}}{\pgfpoint{10pt}{10pt}}
\pgfrect[stroke]{\pgfpoint{40pt}{70pt}}{\pgfpoint{10pt}{10pt}}
\end{pgfpicture}
\end{center}
For the full Calogero system, there are further degeneracies due to the
fact that every $B_n$ excitation is degenerate to a $F_n$
excitation; which is simply a consequence of the manifest
supersymmetry. Thus a state with energy $n$ can, once again be
represented by a Young diagram with $n$ boxes, but each one of the
columns (of length $n_i$) now has the option of corresponding to
either a $B_{n_i}$ or a $F_{n_i}$ excitation. We can denote the
columns corresponding to the $F$ excitations by drawing them with
boxes with crosses as depicted below. Hence the complete set of
degenerate states for the Calogero model, corresponding to an
excitation of energy $n$, are described by first forming all the
Young diagrams corresponding to the partitions of $n$. The action of
the supersymmetry generators can then be described by replacing the
columns with the ones containing crossed boxes, one column at a
time. For example, the effect of replacing two columns with crossed
ones is depicted below.
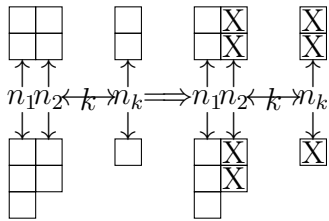
\begin{center}
\begin{pgfpicture}{0cm}{0cm}{4.3cm}{3cm}
\pgfrect[stroke]{\pgfpoint{0cm}{0cm}}{\pgfpoint{10pt}{10pt}}
\pgfrect[stroke]{\pgfpoint{0cm}{10pt}}{\pgfpoint{10pt}{10pt}}
\pgfrect[stroke]{\pgfpoint{0cm}{20pt}}{\pgfpoint{10pt}{10pt}}
\pgfputat{\pgfpoint{5pt}{35pt}}{\pgfbox[center,center]{$\downarrow$}}
\pgfputat{\pgfpoint{5pt}{45pt}}{\pgfbox[center,center]{$n_1$}}
\pgfputat{\pgfpoint{5pt}{55pt}}{\pgfbox[center,center]{$\uparrow$}}
\pgfrect[stroke]{\pgfpoint{0cm}{60pt}}{\pgfpoint{10pt}{10pt}}
\pgfrect[stroke]{\pgfpoint{0cm}{70pt}}{\pgfpoint{10pt}{10pt}}
\pgfrect[stroke]{\pgfpoint{10pt}{10pt}}{\pgfpoint{10pt}{10pt}}
\pgfrect[stroke]{\pgfpoint{10pt}{20pt}}{\pgfpoint{10pt}{10pt}}
\pgfputat{\pgfpoint{15pt}{35pt}}{\pgfbox[center,center]{$\downarrow$}}
\pgfputat{\pgfpoint{15pt}{45pt}}{\pgfbox[center,center]{$n_2$}}
\pgfputat{\pgfpoint{15pt}{55pt}}{\pgfbox[center,center]{$\uparrow$}}
\pgfrect[stroke]{\pgfpoint{10pt}{60pt}}{\pgfpoint{10pt}{10pt}}
\pgfrect[stroke]{\pgfpoint{10pt}{70pt}}{\pgfpoint{10pt}{10pt}}
\pgfputat{\pgfpoint{25pt}{45pt}}{\pgfbox[center,center]{$\leftarrow$}}
\pgfputat{\pgfpoint{30pt}{45pt}}{\pgfbox[center,center]{$k$}}
\pgfputat{\pgfpoint{35pt}{45pt}}{\pgfbox[center,center]{$\rightarrow$}}
\pgfrect[stroke]{\pgfpoint{40pt}{20pt}}{\pgfpoint{10pt}{10pt}}
\pgfputat{\pgfpoint{45pt}{35pt}}{\pgfbox[center,center]{$\downarrow$}}
\pgfputat{\pgfpoint{45pt}{45pt}}{\pgfbox[center,center]{$n_k$}}
\pgfputat{\pgfpoint{45pt}{55pt}}{\pgfbox[center,center]{$\uparrow$}}
\pgfrect[stroke]{\pgfpoint{40pt}{60pt}}{\pgfpoint{10pt}{10pt}}
\pgfrect[stroke]{\pgfpoint{40pt}{70pt}}{\pgfpoint{10pt}{10pt}}
\pgfputat{\pgfpoint{60pt}{45pt}}{\pgfbox[center,center]{$\Longrightarrow$}}
\pgfrect[stroke]{\pgfpoint{70pt}{0cm}}{\pgfpoint{10pt}{10pt}}
\pgfrect[stroke]{\pgfpoint{70pt}{10pt}}{\pgfpoint{10pt}{10pt}}
\pgfrect[stroke]{\pgfpoint{70pt}{20pt}}{\pgfpoint{10pt}{10pt}}
\pgfputat{\pgfpoint{75pt}{35pt}}{\pgfbox[center,center]{$\downarrow$}}
\pgfputat{\pgfpoint{75pt}{45pt}}{\pgfbox[center,center]{$n_1$}}
\pgfputat{\pgfpoint{75pt}{55pt}}{\pgfbox[center,center]{$\uparrow$}}
\pgfrect[stroke]{\pgfpoint{70pt}{60pt}}{\pgfpoint{10pt}{10pt}}
\pgfrect[stroke]{\pgfpoint{70pt}{70pt}}{\pgfpoint{10pt}{10pt}}
\pgfrect[stroke]{\pgfpoint{80pt}{10pt}}{\pgfpoint{10pt}{10pt}}
\pgfrect[stroke]{\pgfpoint{80pt}{20pt}}{\pgfpoint{10pt}{10pt}}
\pgfputat{\pgfpoint{85pt}{35pt}}{\pgfbox[center,center]{$\downarrow$}}
\pgfputat{\pgfpoint{85pt}{45pt}}{\pgfbox[center,center]{$n_2$}}
\pgfputat{\pgfpoint{85pt}{55pt}}{\pgfbox[center,center]{$\uparrow$}}
\pgfrect[stroke]{\pgfpoint{80pt}{60pt}}{\pgfpoint{10pt}{10pt}}
\pgfrect[stroke]{\pgfpoint{80pt}{70pt}}{\pgfpoint{10pt}{10pt}}
\pgfputat{\pgfpoint{85pt}{15pt}}{\pgfbox[center,center]{X}}
\pgfputat{\pgfpoint{85pt}{25pt}}{\pgfbox[center,center]{X}}
\pgfputat{\pgfpoint{85pt}{65pt}}{\pgfbox[center,center]{X}}
\pgfputat{\pgfpoint{85pt}{75pt}}{\pgfbox[center,center]{X}}
\pgfputat{\pgfpoint{95pt}{45pt}}{\pgfbox[center,center]{$\leftarrow$}}
\pgfputat{\pgfpoint{100pt}{45pt}}{\pgfbox[center,center]{$k$}}
\pgfputat{\pgfpoint{105pt}{45pt}}{\pgfbox[center,center]{$\rightarrow$}}
\pgfrect[stroke]{\pgfpoint{110pt}{20pt}}{\pgfpoint{10pt}{10pt}}
\pgfputat{\pgfpoint{115pt}{35pt}}{\pgfbox[center,center]{$\downarrow$}}
\pgfputat{\pgfpoint{115pt}{45pt}}{\pgfbox[center,center]{$n_k$}}
\pgfputat{\pgfpoint{115pt}{55pt}}{\pgfbox[center,center]{$\uparrow$}}
\pgfrect[stroke]{\pgfpoint{110pt}{60pt}}{\pgfpoint{10pt}{10pt}}
\pgfrect[stroke]{\pgfpoint{110pt}{70pt}}{\pgfpoint{10pt}{10pt}}
\pgfputat{\pgfpoint{115pt}{25pt}}{\pgfbox[center,center]{X}}
\pgfputat{\pgfpoint{115pt}{65pt}}{\pgfbox[center,center]{X}}
\pgfputat{\pgfpoint{115pt}{75pt}}{\pgfbox[center,center]{X}}
\end{pgfpicture}
\end{center}
One also has to impose the rule that in any given Young diagram one
can have at the most one 'crossed' column of a given length. This
simply follows from the fact \beq F_{n_i}F_{n_i} =0\eeq
To each such partition, one can associate a state of the Calogero
model. The naive association of states to partitions would be to
associate the the appropriate an oscillator excitation to every
column of the Young diagram, i.e., an excitation of the $B_n$($F_n$)
type for each uncrossed (crossed) column of length $n$. Although the
states so formed would be bona fide eigenstates of the Calogero
Hamiltonian, they will not diagonalize the Higher conserved charges
of the Calogero system. This is the analog of the difference between
the string basis and the basis formed my taking the Slater
determinants of the various Hermite polynomials for the free Fermion
system \cite{beren-2}. However, the eigenstates that diagonialize all
the mutually commuting charges of the Calogero system were
identified in \cite{jack1} and their relation to the partitions
described above was also made clear in the same paper. Since, we
shall not be involved in the diagonalization of the higher charges
in the present work, we shall refer to \cite{jack1} for further
details of the construction of eigenfunctions.

Having enumerated the degeneracies of the Calogero model we can now
proceed to apply these results to the dilatation operator $H'$. The
dilatation operator differs from the Calogero system by a term
proportional to the Fermion number operator. However the above
discussion can be easily generalized to understand its degeneracies
as well. In the matrix model language, the Hamiltonian is
\beq H' =
\tr( A^\dagger A + \frac{3}{2}\Psi ^\dagger \Psi ),\eeq and the
factor of three halves in front of the Fermion number operator is
due to the fact that the dilatation operator measures the conformal
dimensions of the gauge theory composite operators and the Fermions
have a bare conformal dimension of $\frac{3}{2}$ while that for the
scalars os $1$. In the basis, where the position space matrix
corresponding to $A$ is diagonal, the dilatation operator $H +
\frac{1}{2} \mathcal{A}^{\dagger i}\mathcal{A}_i$ is: \beq H' =
\sum_i\left(-\frac{1}{2}\frac{\partial }{\partial x_i^2} +
\frac{3}{2}\mathcal{A}^{\dagger i}\mathcal{A}_i +
\frac{1}{2}x_i^2\right) + \frac{1}{2} \sum_{i\neq j}\left(
\frac{\mathcal{A}^{\dagger i}\mathcal{A}_i - \mathcal{A}^{\dagger
i}\mathcal{A}_j}{(x_i - x_j)^2}\right).\eeq
The previous discussion
about states being labeled by $F$ and $B$ type oscillators goes
through but the degeneracies are to be counted in a somewhat
different manner. From the Hamiltonian, it is clear that three
Bosonic excitations have the same energy as two Fermionic ones, thus
$B_n$ and $F_n$ no longer represent degenerate excitations. However
$B_n$ and $F_{n_1}F_{n-n_1-1}$ do, for every value of $n_1$. Thus,
as before, the degeneracies can be counted using  Young diagrams.
For a given excitation of energy $n$ one again forms all the Young
diagrams corresponding to the partitions of $n$. These are simply
all the zero Fermion number excitations. One can then replace each
column of the Young diagram (say of length $m$) with two crossed
columns of lengths $m_1$ and $m_2$ satisfying\beq m_1 +m_2 =
m-1.\eeq The new columns have to be added in a way such that the new
diagram is still a legal Young diagram. Each such replacement is
equivalent to replacing three Bosonic excitations with two Fermionic
ones. Carrying this process out for all the columns of the diagrams
generates for us all the $F$ type excitations that are degenerate to
a state of a given energy. In the process of generating Fermionic
excitations, one also needs to exercise the constraint that there
cannot be two crossed columns of the same length in a given Young
diagram.

The effect of replacing Bosonic excitations by Fermionic ones of
length 1 on a particular young diagram is illustrated in the
following diagram.
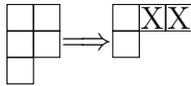
\begin{center}
\begin{pgfpicture}{0cm}{0cm}{2.5cm}{1.5cm}
\pgfrect[stroke]{\pgfpoint{0cm}{0cm}}{\pgfpoint{10pt}{10pt}}
\pgfrect[stroke]{\pgfpoint{0cm}{10pt}}{\pgfpoint{10pt}{10pt}}
\pgfrect[stroke]{\pgfpoint{0cm}{20pt}}{\pgfpoint{10pt}{10pt}}
\pgfrect[stroke]{\pgfpoint{10pt}{20pt}}{\pgfpoint{10pt}{10pt}}
\pgfrect[stroke]{\pgfpoint{10pt}{10pt}}{\pgfpoint{10pt}{10pt}}
\pgfputat{\pgfpoint{30pt}{15pt}}{\pgfbox[center,center]{$\Longrightarrow
$}} \pgfrect[stroke]{\pgfpoint{40pt}{20pt}}{\pgfpoint{10pt}{10pt}}
\pgfrect[stroke]{\pgfpoint{40pt}{10pt}}{\pgfpoint{10pt}{10pt}}
\pgfrect[stroke]{\pgfpoint{50pt}{20pt}}{\pgfpoint{10pt}{10pt}}
\pgfrect[stroke]{\pgfpoint{60pt}{20pt}}{\pgfpoint{10pt}{10pt}}
\pgfputat{\pgfpoint{55pt}{25pt}}{\pgfbox[center,center]{X}}
\pgfputat{\pgfpoint{65pt}{25pt}}{\pgfbox[center,center]{X}}
\end{pgfpicture}
\end{center}
In the usual analysis of Calogero systems with a finite number ($N$)
of particles, one imposes a non-perturbative cutoff on the depth of
the columns of the Young diagrams. Namely, the columns are not
allowed to have more than $N$ boxes. However, that would correspond
to the finite $N$ matrix model, for which the states that we picked
are no longer protected. The large $N$ limit, translates, in the
language of the Young diagrams to lifting the non-perturbative
cutoff on the depth of the columns. Looked at in another way,
imposing the BPS condition at the level of the Calogero system is
equivalent to lifting the  cutoff on the depth of the columns.

\section{Remnants of Yangian Symmetries and Loop Algebras:}In this
section, we shall focus on the realization of Yangian symmetries and
non-local conservation laws in the super-Calogero system.

The Calogero model is nothing but the gauge fixed form of the
dilatation operator in a particular sector of BPS operators. If one
goes beyond the BPS sectors, one would of course have to incorporate
the perturbative corrections to the dilatation generator. At finite
values of $N$ this is hard to accomplish, however from the detailed
studies of operator mixing in the gauge theory in the recent past,
the first few perturbative corrections to the large $N$ limit of the
dilatation operator are known in rather explicit forms, at least in
some small sectors of operator mixing. For example, in the $su(2|3)$
sector discussed earlier, the planar dilatation operator is known up
to the third order in the 't Hooft coupling \cite{dyn}. It has also
been shown that the dilatation operator can be realized as an
integrable quantum spin chain up to this order in perturbation
theory \cite{dyn, ab-dyn}. One point of view on the integrability of
the spin chain relates the integrability to the existence of Hopf
algebraic symmetries: the integrability being simply the
manifestation of such large hidden symmetries.  For more detailed
studies of the Yangian for the gauge theory we shall refer to
\cite{gauge-yan-1,gauge-yan-2,gauge-yan-3}. The existence of Yangian
symmetries, apart from providing key insights into the algebraic
structures that are responsible for the integrability of the spin
chain, are also crucial from the point of view of the AdS/CFT
correspondence as the string sigma model has been known to possess
this very same symmetry at the classical level \cite{string-yan-1,
string-yan-2,string-yan-3,string-yan-4,string-yan-5}. To the extent
that the spectrum of anomalous dimensions of the gauge theory and
those closed string excitations agree (for instance in the BMN
limit) it has been possible to relate the Yangian symmetries on the
gauge theory and the gravity sides. For the specific case of studies
of integrable structures in the the $su(1|1)$ sector of the AdS/CFT
correspondence we shall refer to \cite{gauge-yan-2,string11-1,
string11-2}

However, it is not clear at the moment whether or not these novel
symmetries survive the low energy supergravity limit. One can
however use the gauge theory as a probe to investigate this problem.
Since results from the BPS sectors of the gauge theory can be
extrapolated to the supergravity limit one can investigate the role
of the Yangian symmetries of the dilatation operator when it is
restricted to the BPS states and try and understand how these
symmetries manifest themselves in the supergravity limit. With this
motivation in mind we can probe the structure of Yangian symmetries
of the super Calogero model studied so far.

The Yangian charges and the conserved integrals of motion of the
Calogero model are generated by the matrix elements of the transfer
matrix, which is a $2\times 2 $ matrix for the $su(1|1)$ model, each
matrix element of which is an operator in the Hilbert space of the
Calogero model \cite{yan-koo}.\footnote{There is a large literature
on the role of Yangian symmetries and quantum spin chains. Of
particular relevance to the present problem are \cite{bernard-rev,
ber-dressing, ber-long-rng-1, haldane-1, hal-charge}.}
 The transfer matrix has a free parameter, namely the spectral
parameter $u$ and the standard expansion around an infinite vale of
the spectral parameter reads as

\beq T^{ab} = I\delta ^{ab} + \sum_{n=1}^{\infty }
\frac{1}{u^n}T_{n-1}^{ab}.\eeq

In the above expression $S^{ab}(j)$ are the $su(1|1)$ generators at
the $j$ th lattice site,\beqs S^{11}(j) =
\mathcal{A}_{j}\mathcal{A}^\dagger _{j}, S^{22}(j) =
\mathcal{A}^\dagger _{j}\mathcal{A} _{j}\nonumber \\
S^{12}(j) =\mathcal{A} _{j}, S^{21}(j) =\mathcal{A}^\dagger
_{j}.\eeqs

\beq T_{n}^{ab} = \sum _{j,k} S^{ab}(j)(L^n)_{j,k},\eeq where $L^n$
is the $n$ th power of the $N\times N$ Lax matrix
 \beq L_{j,k} = \delta _{j,k}(\frac{\partial }{\partial x_j}+x_j) + \hbar
(1-\delta _{j,k}) \omega _{j,k}\Pi_{j,k}\eeq and \beq \omega _{j,k}
= \frac{e^{-\frac{\hbar}{2}(x_i - x_j)}}{\sinh \frac{\hbar}{2}(x_i -
x_j)}\eeq We have chosen to incorporate a free parameter, which we
suggestively denote by  $\hbar $ in the above analysis to illustrate
the contraction of the Yangian algebra to the loop algebra in a
transparent way. We have also chosen an inverse hyperbolic fall off
of the inter-particle potential in the Lax operator rather than the
$1/(x_i - x_j)$ fall off for the same purpose. The basic idea being
to start with the hyperbolic case, which contains the rational and
the trigonometric Calogero models as special cases and recover the
underlying symmetry of the rational case as a particular limit.

The transfer matrix satisfies the quadratic Yang-Baxter algebra.

\beq [T^{ab}_s, T^{cd}_{p+1}]_\pm - [T^{ab}_{p+1}, T^{cd}_{s}]_\pm =
\hbar (-1)^{\epsilon (c) \epsilon (a) + \epsilon (c) \epsilon (b) +
\epsilon (b) \epsilon (a)}\left( T^{cb}_{p} T^{ad}_{s} - T^{cb}_{s}
T^{ad}_{p}\right).\eeq In the above equation $\epsilon $ denotes the
grade $\epsilon (1) = 0, \epsilon (1) = 1$. It is important to note
that the non-linearity of the Yang-Baxter algebra (the r.h.s of the
above equation) is proportional to $\hbar $. The Yang-Baxter algebra
also implies \beq \sum_{i,j}[T^{ii}_m, T^{jj}_n] = 0\eeq i.e the
trace of the transfer matrix is the generating function for the
conserved charges which are in involution. As a matter of fact, if
one denoted these charges by $H_n = \tr T^n$, then one can show that
the Hamiltonian, up to the addition of constant terms is nothing but
$T^2$, which for the Hyperbolic case takes on the following form.

 \beq H =
\frac{1}{2}\sum_{j,k}(-\partial ^2_j + x_j^2 + \mathcal{A} ^\dagger
(j)\mathcal{A} (j) + \hbar \Pi _{j,k}\partial _j \omega _{j,k} +
\hbar ^2\omega _{j,k}\omega _{k,j})\eeq

For the limit of interest to us, $\hbar \rightarrow 0 $ we recover
\beq \omega _{j,k} = \frac{1}{x_j - x_k}\eeq with the Hamiltonian
above becoming the super-Calogero Hamiltonian :\beq H \rightarrow
\sum_i\left(-\frac{1}{2}\frac{\partial }{\partial x_i^2} +
2\mathcal{A}^{\dagger i}\mathcal{A}_i + \frac{1}{2}x_i^2\right) +
\frac{1}{2}\sum_{i\neq j}\left( \frac{1-\Pi_{i,j}}{(x_i -
x_j)^2}\right).\eeq As is obvious from the above construction,  in
this 'clasical' limit, the Yangian algebra degenerates into the loop
algebra: \beq [T^{ab}_s, T^{cd}_{p+1}]_\pm - [T^{ab}_{p+1},
T^{cd}_{s}]_\pm = 0,\eeq which can be written, upon using the above
relations recursively as: \beq [T^{ab}_s, T^{cd}_{p}]_\pm =
\delta_{b,c} T^{ad}_{p+s} - (-1)^{(\epsilon (a) + \epsilon
(b))(\epsilon (c) + \epsilon (d)}\delta_{a,d} T^{cb}_{p+s}\eeq
Hence, the integrable structure in the dynamics of
the $su(1|1)$ BPS operators appears to arise from the loop algebra of
$su(1|1)$. One might have anticipated this from the fact that
the dynamics of the BPS sectors of the gauge theory can be
extrapolated to the supergravity regime and the supergravity can be
regarded as a classical limit of the string theory. On the other
hand the loop algebra is also a classical limit of the Yangian
algebra, which appears to be a symmetry of the dual string theory.
The discussion above indicates, through an explicit construction, that
these two notions of classical limits are compatible with each
other.

Furthermore, we can also see that the supersymmetry generators are
contained in the loop algebra. As a matter of fact it is easy
to see that: \beq T^{21}_1 = Q, T^{12}_1 = Q^\dagger \eeq and
that:\beq H = [ T^{21}_1, T^{12}_1]_+\eeq

The higher (odd) elements of the loop algebra simply act as the
supersymmetry generators for the higher conserved charges of the
system:\beq H_{n+m} = [ T^{12}_n, T^{21}_m]_+ \eeq We thus see that
the loop algebra and the supersymmetry of the particle mechanics fit
together in a natural way.

The Calogero model Hamiltonian is, of course,
not the dilatation operator, as the
two differ by a term proportional to the fermion number operator
\beq H' = H +
\frac{1}{2}\mathcal{A}^\dagger_i\mathcal{A}_i\eeq However, just as
we were able to recover the degeneracies of the dilatation operator
from those of the Calogero model, it is possible to use the transfer
matrix of the Calogero system to construct the integrals of motion
for the dilatation operator. The construction is extremely simple.
The Fermion number operator does not commute with the supersymmetry
generators, and in general, with the odd elements of the Yangian
algebra $T^{12}_n $ and $T^{21}_m$. However, it does commute with
the generators of even grade. Thus we have \beq [H', T^{ii}_n] = 0
~~ \forall i\eeq
Moreover, from the loop algebra it is clear that \beq
[T^{11}_m, T^{11}_n] = [T^{11}_m, T^{22}_n] = [T^{22}_m, T^{22}_n] =
0 ~~\forall m,n\eeq Thus we have as many conserved charges in
involution for the dilatation operator as there are degrees of
freedom; namely $2N$. Thus we recover the integrability of the
dilatation operator from the underlying loop algebraic symmetry of
the super Calogero model.

\section{Spectrum of the Euler-Calogero System}
We shall now revert back to the
general $su(2|3)$ Hamiltonian given in (\ref{euler-calogero}).
Integrability of the particle mechanics model presented above
derives from the fact that it is nothing but a sum of decoupled
(matrix) oscillators in disguise. Such matrix models are obviously
integrable, indeed even for finite values of $N$, and they continue to
be integrable in the large $N$ limit. Apart from the explicit
solutions to the equations of motion of these matrix models,
integrability also manifests in the existence of
a large number (infinite in the large $N$ limit) of conserved
quantities. It is worthwhile to understand the integrability of the
many-body system in some detail. To do that let us begin by writing
the Hamiltonian in the special basis in which $X^1$ is diagonal:
\beq
H = \sum_\alpha (a^{\dagger \alpha} a_\alpha).\eeq It is understood
that\beqs(a_1)^i_j = \left( x_i + \frac{\partial}{\partial
x_i}\right)\delta^i_j + \frac{(1-\delta
^i_j)\mathcal{L}^i_j}{x_i-x_j}\nonumber
\\(a^{\dagger 1})^i_j = \left( x_i - \frac{\partial}{\partial
x_i}\right)\delta^i_j + \frac{(1-\delta
^i_j)\mathcal{L}^i_j}{x_i-x_j}\label{lax}\eeqs and the other
oscillators $a^{\dagger \alpha}, a_\alpha$ ($\alpha \neq 1$) are
simply the remaining degrees of freedom for the gauged fixed matrix
model i.e they are the $U(N)$ rotated oscillators. Translating the
original matrix model equations of motion to this special basis one
can see that\beqs \dot{a}^{\dagger \alpha} = a^{\dagger \alpha} +
[a^{\dagger \alpha}, g]\nonumber \\ \dot{a}_{\alpha} = -a_{\alpha} +
[a_{\alpha}, g]\eeqs where the commutator on the r.h.s is the matrix
commutator and \beq g^i_j = \frac{(1-\delta
^i_j)\mathcal{L}^i_j}{x_i-x_j}\eeq It is now a straightforward
exercise to show that operators\beq \mathcal(O)^{\alpha_1\cdots
\alpha _m}_{\beta _1 \cdots \beta _n} = \tr\left(a^{\dagger \alpha
_1}\cdots a^{\dagger \alpha_m}a_{\beta _1}\cdots a_{\beta
_n}\right)\eeq evolve according to\beq
\dot{\mathcal{O}}^{\alpha_1\cdots \alpha _m}_{\beta _1 \cdots \beta
_n} = (m-n)\mathcal(O)^{\alpha_1\cdots \alpha _m}_{\beta _1 \cdots
\beta _n}\eeq This obviously implies that
$\mathcal(O)^{\alpha_1\cdots \alpha _n}_{\beta _1 \cdots \beta _n}$
are all integrals of motion for every $n$ and that the states\beq
|\{\alpha _{i_1} \cdots \alpha _{i_m}\}> = \tr a^{\dagger \alpha
_1}\cdots a^{\dagger \alpha _m}|0>\eeq are exact eigenstates of $H$
with energy $m$. Thus, quite like the super-Calogero model the
degeneracies can once again be counted by the use of Young diagrams.
In the zero Fermion number sector, $\mathcal{L}^i_j =0$, and hence,
all the states with a given energy $n$ can be labeled by Young
diagrams corresponding to the partitions of $n$. But unlike the
Calogero model, one now has four types of impurities, two Bosonic
and two Fermionic. Just as we introduced diagrams with crossed
columns in the Calogero case, here we have to distinguish between
the various impurities, and hence it is useful to think of the
columns being colored by four colors corresponding to the
impurities. Thus the additional degeneracies are generated by
replacing the columns of the Young diagrams of the zero impurity
number sector with colored columns one at a time. We also have to
keep in mind that when we add columns corresponding to the Bosonic
impurities, we trade a column of the original free Fermion Young
diagram for a colored column of the same length.  However, as in the
case of the $su(1|1)$ sector, when it comes to inserting Fermionic
impurities, one has to replace the columns of the  free Fermion
Young diagram (say of size $n$) with two columns, one of lengths
$n_1$ and $n_2$ satisfying\beq n_1 +n_2 = n-1.\eeq Furthermore,  we
need to make sure that the there is at the most one Fermionic column
of a given color and length.

This construction counts all the degeneracies between states that
have at the most one impurity inside a single trace in the original
matrix model picture. These are half BPS states, although not all
states of the matrix model are. The degeneracies between states of
zero impurity number  and impurity number greater than one are not
accounted for by the above construction. For example, the above
construction does not count the degeneracy between the states
\beq
\tr(a^{1\dagger})^{m+n+2} |0> ~~\mbox{and}~~
\tr\left((a^{1\dagger})^{m}a^{2\dagger}(a^{1\dagger})^{n}a^{2\dagger}
\right)|0>\nonumber.\eeq The second state above is non-BPS. Thus, we
have been able to utilize the integrability of the Euler-Calogero
model to enumerate all the BPS states, that are charged under
$su(2|3)$, formed out of inserting a single impurity field inside a
matrix trace.

A natural question that arises is how one may describe BPS
excitations involving several impurity fields within the open string
picture. The answer to that is not hard to see. One needs to write
the supercharges for the full $su(2|3)$ sector in the basis in the
which $X^1$ is diagonal. Since the $BPS$ states are all generated by
the action of the supersymmetry generators, all one needs to do is
write the super charges in this basis and generate all the  $BPS $
states by their repeated action. The supercharge of interest to us
is the one that replaces a scalar impurity field by a Fermionic one
and it can be written as $2\times 3$ matrix, with matrix
elements
\beq Q^I_i = \tr(\Psi ^{\dagger I}a_i)\eeq
which in the basis
of interest takes the form \beq Q^\alpha_\beta = \tr(a^{\dagger
\alpha}a_\beta), ~~\alpha = 4,5, ~\beta = 1,2,3.\eeq
Needless to say,
in this second form it is implied that the oscillators
are the ones in the $U(N)$ rotated basis (\ref{lax}).

The construction described previously enumerates all the BPS states
formed out of single action of the supercharge. The rest can be
similarly generated by repeating the action of the supersymmetry
generator given above. This will pick out all the $BPS$ states
contained in the complete set of states of the Euler-Calogero model.

\section{Discussion and Future Directions}
The general connection between the dilatation operator and Calogero
systems can lead to  several interesting avenues of investigation
that were not addressed in the present work. We list some of these
possibilities below.
\\
{\bf 1:} From the point of view of integrable systems, it would be
extremely interesting to study the integrability of the
Euler-Calogero system in greater detail. In the present work, we
presented enough of an understanding of its integrability to
understand its spectrum and the associated degeneracies. Gaining an
understanding of the underlying Yang-Baxter algebra for the quantum
Euler-Calogero system would clarify the role of Yangian type
symmetries for this system. Such an analysis should be possible, as
the classical $r$ matrix for the Euler-Calogero system, which
curiously enough is a dynamical `r' matrix has been found
in \cite{eul-r}. Of particular interest would be the a systematic
understanding of the dynamical models and integrable structures that
arise when one considers BPS operators that involve more than a
single impurity field inside a trace. \\
{\bf 2:} In the paper we showed that the rational super-Calogero
model can be regarded as the simplest non-trivial generalization of
the theory of free Fermions when it comes to understanding protected
operators of the gauge theory. Just like the theory of free
Fermions, it was shown that one can have two equivalent description
of the states of this theory, which we regarded as an open/closed
duality. Clearly it would be extremely desirable to have a world
sheet interpretation of the super-Calogero system, along the lines
of the description provided in \cite{ho-string} for the free Fermion
system. It is not hard to envisage what the world sheet string
theory would be. The string dual of the free Fermion system was
found by taking valuable clues from string theory in two dimensions
and analytically continuing the string dual of the $C=1$ matrix
model to the case of the `right-side-up' harmonic oscillator. To
take a similar clue for the string dual of the Calogero model, we
shall have to look at the world-sheet description of strings in
$AdS_2$. This particular string theory was analyzed recently
in \cite{her-mar} and a connection to Calogero systems was also made
in the same paper. It seems plausible that this very theory is the
string dual of the $su(1|1)$ BPS sector of $\mathcal{N}$ =4 SYM
discussed earlier in this paper. We hope to report on this
possibility in the near future.\\
{\bf3:} Clearly, the tree level dilatation operator can be written
as an Euler-Claogero system even if the states in question do not
correspond to BPS operators of the gauge theory. Hence the
Euler-Calogero system provides us with a starting point for
understanding non-BPS excitations. It would indeed be extremely
interesting to understand how this framework of the Euler-Calogero
model changes once the higher loop corrections to the dilatation
operator are considered. Recently it has been shown that it is
possible to obtain the all-loop BMN formula by doing a one loop
computation around a carefully chosen vacuum of the dilatation
operator \cite{all-loop-bmn-1, all-loop-bmn-2}. This point of view
can be easily incorporated within the formalism developed in the
present paper. We hope to report on the connection of Euler-Calogero
type of dynamical systems and non-BPS corrections to the
supergravity spectrum in the near future as well.

\vskip 0.3cm
\noindent
{\bf Acknowledgements:} It is a pleasure to thank Niklas
Beisert, David Berenstein, Avinash Dhar, Dimitra Karaballi, Gautam
Mandal, Parameswaran Nair and Sarada Rajeev for valuable discussions. A.A.\ is
also indebted to the Tata Institute for Fundamental Research for the
hospitality during his stay where a part of this paper was written. The
research of A.P.\ is supported by National Science Foundation grant PHY-0353301
and by PSC-CUNY grant 67526-0036.

\end{document}